\documentclass[prl,twocolumn,amsmath,showpacs]{revtex4}
\usepackage{graphicx}

\begin{document}

\title{Distribution of the superconducting gap in an
YNi$_2$B$_2$C film studied by point contact spectroscopy.}

\author{D.L. Bashlakov, Yu.G. Naidyuk, I.K. Yanson}

\affiliation{B. Verkin Institute for Low Temperature Physics and
Engineering, National Academy of Sciences of Ukraine, 47 Lenin
Ave., 61103 Kharkiv, Ukraine}

\author{S.C. Wimbush \footnote{Present address: National Institute for Materials Science,
International Center for Young Scientists, 1-1 Namiki, Tsukuba,
Ibaraki 305-0044, Japan}, B. Holzapfel, G. Fuchs, S.-L. Drechsler}

\affiliation{IFW Dresden, P. O. Box 270016, 01171 Dresden,
Germany}

\date{\today}

\begin{abstract}
The differential resistances $R_{d}=dV/dI(V)$ of point contacts
between a normal metal and a {\it c} axis oriented
YNi$_{2}$B$_{2}$C film ($T_{\rm c}$ = 15.2\,K) in the
superconducting (SC) state have been investigated. $R_{d}(V)$
contains clear "gap" features connected with processes of Andreev
reflection at the boundary between normal metal and superconductor
that allow the determination of the SC gap $\Delta$ and its
temperature and magnetic field dependence. A distribution of
$\Delta$ from $\Delta_{\rm min}\approx$ 1.5\,meV to $\Delta_{\rm
max}\approx$ 2.4\,meV is revealed; however the critical
temperature $T_{\rm c}$ in all cases corresponded to that of the
film. The value 2$\Delta_{\rm max}$/$k_{\rm B}T_{\rm c}\approx$
3.66 is close to the BCS value of 3.52, and the temperature
dependence $\Delta(T)$ is BCS-like, irrespective of the actual
$\Delta$ value. It is supposed that the distribution of $\Delta$
can be attributed to a gap anisotropy or to a multiband nature of
the SC state in YNi$_{2}$B$_{2}$C, rather than to the presence of
nodes in the gap.

\pacs{74.70.Dd, 74.45.+c, 74.50.+r}
\end{abstract}

\maketitle

\section{Introduction}
During the last ten years, the class of nickel borocarbide
compounds {\it R}Ni$_{2}$B$_{2}$C, where {\it R} = Sc, Y, Lu, Dy,
Ho, Er, Tm and other rare-earth ions, has attracted considerable
attention in the superconducting community \cite{Muller}. This
interest is caused by the unusual properties of these materials
related to the rich interplay of superconductivity and magnetism,
when {\it R} is a magnetic rare-earth ion. In this context, it is
remarkable that the superconducting transition temperature $T_{\rm
c}$ can be higher than the N\'{e}el temperature $T_{\rm N}$ as for
{\it R}=Tm and Er, but also comparable with $T_{\rm N}$ as for
{\it R}=Ho and even lower than $T_{\rm N}$ as for {\it R}=Dy.
Non-magnetic compounds with {\it R}=Sc, Y and Lu have the highest
$T_{\rm c}$ values of 15\,K, 15.5\,K and 16.5\,K, respectively.
They are type II superconductors and have three-dimensional,
practically isotropic electronic structure, while their crystal
structure comprises alternating {\it R}C and N$_{2}$B$_{2}$
layers. Noteworthy recent experiments including tunnelling
\cite{Martinez} and point contact measurements
\cite{Pratap,Bobrov} on bulk single crystals have shown that the
unusual superconducting properties of nickel borocarbides exhibit
strong anisotropy. In particular, a hybrid $(s + g)$ wave
superconducting state has been proposed \cite{Maki1,Maki2} for the
compounds with {\it R} = Y and Lu, characterized by four point
nodes of the superconducting order parameter in the [$\pm $1,0,0]
and [0,$\pm $1,0] directions. This conclusion was based on
investigations of the angular dependence of the thermal
conductivity \cite{Izawa} and specific heat \cite{Park} under an
applied external magnetic field of single crystals at low
temperatures. Point-contact measurements \cite{Pratap} also
revealed a strong gap anisotropy. In the [100] direction, not only
the gap is 4.5 times smaller, but also the transition temperature
was found to be half (!) that in the [001] direction. It is also
important to note that a thorough analysis of upper critical field
$H_{c2}(T)$ data for YNi$_{2}$B$_{2}$C and LuNi$_{2}$B$_{2}$C
single crystals clearly showed that superconductivity in these
compounds is strongly affected by their multiband nature
\cite{Shulga}. In order to gain further insight into the interplay
of the symmetry of superconducting order parameters, their
anisotropy and multiband nature, additional studies are necessary.
In the present work, direct measurements of the gap amplitude
distribution have been performed on epitaxial {\it c} axis
oriented YNi$_{2}$B$_{2}$C thin film using point-contact
spectroscopy \cite{PCSbook}.

\section{Experimental Technique}

In this paper, we present the results of point contact
measurements made on a high quality YNi$_{2}$B$_{2}$C film
\cite{Wimbush}. The film was prepared by pulsed laser deposition
on an MgO single crystal substrate. The SC transition temperature
$T_{\rm c}$ = 15.2\,K (with transition width $\Delta T_{\rm c}=$
0.3\,K) was determined from resistive measurements. The residual
resistance ratio $R(300K)/R(T\geq T_{\rm c})$ of the film was
about 8. X-ray data indicate that the film consists of small
crystallites having their primary orientation with the {\it c}
axis perpendicular to the substrate surface.

The measurement cell was placed in a flow cryostat, enabling
measurement at a series of temperatures up to $T_{\rm c}$ and
higher. Point contacts were created by touching sharpened Cu and
Ag wires to the film surface directly within the cryostat at
liquid helium temperature. An unfortunate disadvantage of this
'needle--anvil' method is the sensitivity of the contacts to
mechanical vibrations and changes in temperature. As a result,
only around one quarter of the prepared contacts withstood
measurement above liquid helium temperature

Using a technique of synchronous detection of weak (10-30\,$\mu$V)
modulating signal harmonics, the first harmonic of the modulating
signal $V_1$ [proportional to the differential resistance
$R_{d}(V)=dV/dI(V)$] was recorded as a function of the bias
voltage $V$.

\begin{figure}
\begin{center}
\includegraphics[width=8cm,angle=0]{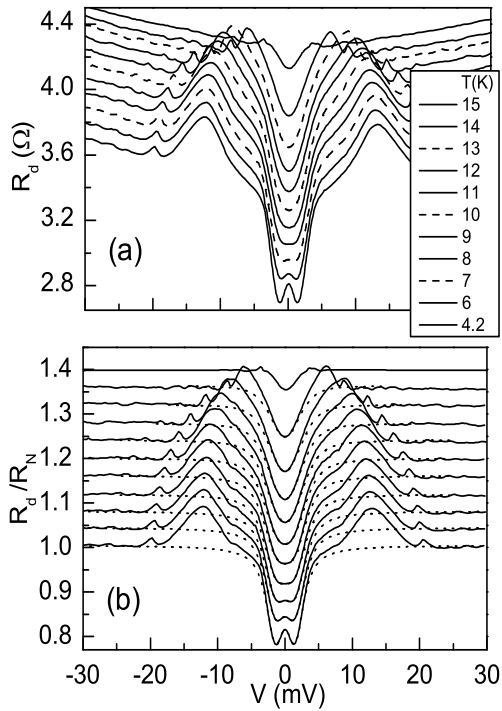}
\end{center}
\caption[] {(a) Differential resistance $R_{d}(V)$ of point
contact YNi$_{2}$B$_{2}$C--Ag with normal state resistance
$R_{N}=3.5 \Omega$, measured at different temperatures. (b)
Normalized $R_{d}$ (solid) curves for this contact and those
values calculated by BTK theory (dashed). The theoretical curve at
4.2\,K is calculated with $\Delta \approx 1.67$\,meV, $Z \approx
0.43$ and $\Gamma \approx 0.55$\,meV. The temperature dependence
$\Delta(T)$ is shown in Fig.\,\ref{yf3}(b). Curves are offset
vertically from the lower curve for clarity.} \label{yf1}
\end{figure}

\section{Experimental Results}

\begin{figure}
\begin{center}
\includegraphics[width=8cm,angle=0]{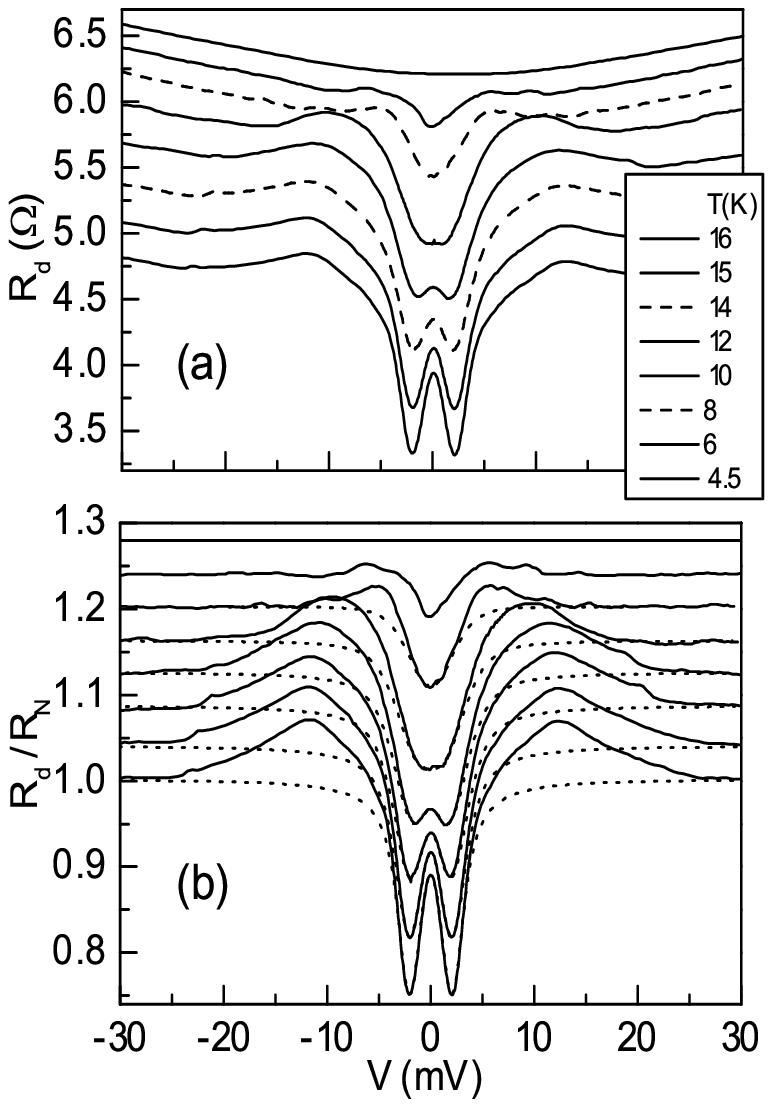}
\end{center}
\caption[] {(a) Differential resistance $R_{d}(V)$ of point
contact YNi$_{2}$B$_{2}$C--Cu with normal state resistance
$R_{N}=4.5 \Omega $, measured at different temperatures. (b)
Normalized $R_{d}$ (solid) curves for this contact and those
values calculated by BTK theory (dashed). The theoretical curve at
4.5\,K is calculated with $\Delta \approx 2.24$\,meV, $Z \approx
0.5$ and $\Gamma \approx 0.1$\,meV. The temperature dependence
$\Delta(T)$ is shown in Fig.\,\ref{yf3}(b). Curves are offset
vertically from the lower curve for clarity.} \label{yf2}
\end{figure}

A series of $R_{d}(V)$ measurements with clear double minimum features near $V=0$
connected with the Andreev reflection of electrons at the superconductor
-- normal metal (S--N) interface are shown in Figs.\,\ref{yf1}(a) and
\ref{yf2}(a).

For voltages exceeding the gap value, all curves exhibit maxima
caused, most likely, by the suppression of superconductivity in
(part of) the contact. Such behavior testifies to a deviation from
the spectral (ballistic) regime (i.e., the inelastic electron mean
free path becomes less than or about the size of the contact) with
increasing bias voltage at the contact (see \cite{PCSbook},
Chapter 3).

The size of the gap was calculated using the modified BTK theory
(see \cite{PCSbook}, Chapter 3.7). The presence of maxima in
$R_{d}(V)$ prevented the coincidence of the calculated and
measured curves throughout the voltage range; therefore an
adjustment was carried out so that both curves coincided in the
region of the gap ($-\Delta$, $+\Delta$) and at the higher
voltages beyond the maxima. Preliminary, unadjusted curves were
normalized to $R_{d}(V,~T>T_{\rm c})$. During calculation, the
parameters $Z$ and $\Gamma $ were fixed or varied in small limits
up to 5--10\,{\%}. The results are shown in Figs.\,\ref{yf1}(b)
and \ref{yf2}(b). We note also that the contacts shown in
Figs.\,\ref{yf1} and \ref{yf2} had, respectively, the minimal and
maximal gap values among all the contacts that withstood low
temperature measurement.

\begin{figure}
\begin{center}
\includegraphics[width=8cm,angle=0]{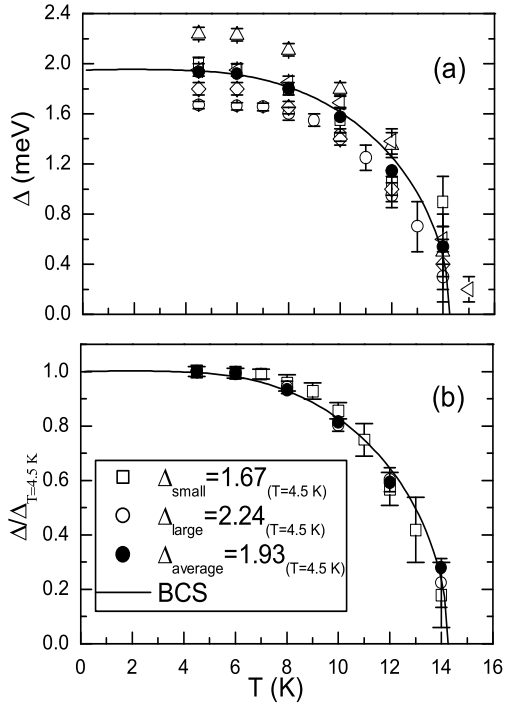}
\end{center}
\caption[] { (a) The temperature dependence of the gap $\Delta(T)$ obtained from
the BTK fit for five different contacts (open symbols). The theoretical BCS
curve is drawn to coincide with the average
value of the five contacts (closed symbols). (b)  The temperature
dependence $\Delta(T)$ normalized to $\Delta(T=$ 4.5\,K) for
minimal (open squares), maximal (open circles) and average (closed
circles) gap values, along with the theoretical BCS curve.} \label{yf3}
\end{figure}

In Fig.\,\ref{yf3}(a), the dependences $\Delta(T)$ are shown for a
number of different contacts, for each of which a series of
$R_{d}(V) $ curves at different temperatures was measured. For all
these contacts, the SC minimum in $R_{d}(V)$ disappeared at a
temperature interval of 14--16\,K, corresponding to the $T_{\rm
c}$ value (15.2\,K) of the film.

In Fig.\,\ref{yf3}(b), the temperature dependences for the maximal and
minimal $\Delta(T=4.5$ K) are shown, normalized to these values (2.24\,meV and
1.67\,meV, respectively). Also the normalized $\Delta(T)$
temperature dependence obtained by averaging all five contacts of
Fig.\,\ref{yf3}(a), and the theoretical BCS dependence are depicted.
We reiterate that the complete data (measured, normalized and calculated
spectra) for the smallest and largest gap are presented in Figs.\,\ref{yf1} and\,\ref{yf2}.

\begin{figure}
\begin{center}
\includegraphics[width=7cm,angle=0]{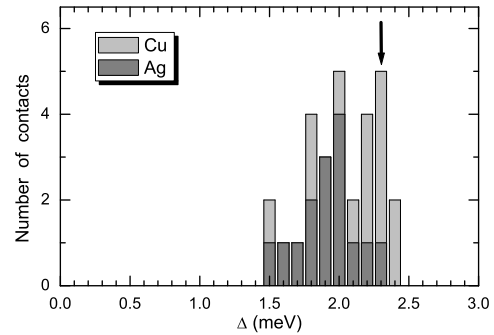}
\end{center}
\caption[] {Distribution of $\Delta$ values obtained from
measurement of 29 contacts with different counter electrodes Ag
(15 contacts, dark bars) and Cu (14 contacts, gray bars). The
arrow specifies the value corresponding to the BCS ratio
2$\Delta_{0}/k_{\rm B}T_{\rm c}=3.52$, with $T_{\rm c}=15.2$\,K. }
\label{yf4}
\end{figure}

In total, 29 curves, on which the structure with two minima is
clear enough to allow an unequivocal determination of $\Delta$
using the standard procedure of adjustment of theoretical curves
described above, have been selected from numerous measured
spectra. From the results of the calculation, the histogram in
Fig.\,\ref{yf4} was constructed. Irrespective of the counter
electrode used, $\Delta$ is distributed in an interval between
1.5\,meV and 2.4\,meV.

\begin{figure}
\begin{center}
\includegraphics[width=9cm,angle=0]{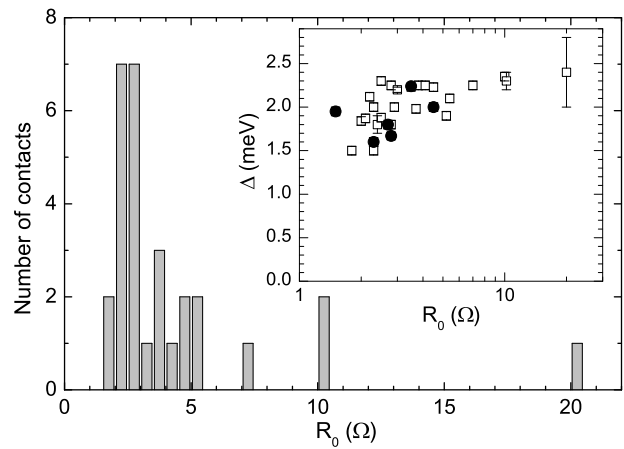}
\end{center}
\caption[] {Distribution of contact resistances $R_0=R_{d}(V=0$,
$T=4.2$\,K) for the same contacts as in Fig.\,\ref{yf4}. Inset:
gap values of these contacts. Solid circles designate those
contacts for which a series of temperature measurements were
carried out, with results shown on Fig.\,\ref{yf3}(a).}
\label{yf5}
\end{figure}

Half of the selected contacts had contact resistances between
2\,$\Omega$ and 3\,$\Omega$, as seen on the distribution of $N(R)$
values in Fig.\,\ref{yf5}. The inset shows the $\Delta$ values
obtained for contacts of different normal state resistance. Solid
circles designate those contacts for which a series of temperature
measurements were carried out and superconductivity confirmed to
disappear between 15\,K and 16\,K, corresponding to the $T_{\rm
c}$ value of the film, independent of the $\Delta$ value.

\begin{figure}
\includegraphics[width=9cm,angle=0]{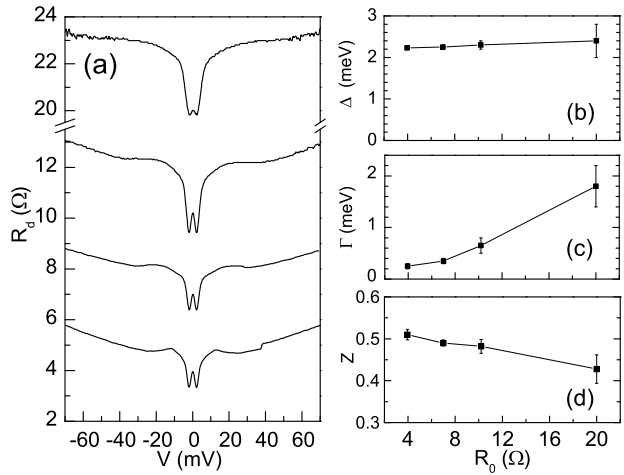}
\caption[] {(a) Differential resistance $R_{d}(V)$ of the same
point contact YNi$_{2}$B$_{2}$C--Cu measured for different
$R_0=R_{d}(V=0$, $T=4.2$\,K) values as a result of consecutive
contact short-circuits by a current pulse. (b), (c), (d)  The
variation in the values $\Delta$, $\Gamma$ and $Z$ with $R_0$.}
\label{yf6}
\end{figure}

In Fig.\,\ref{yf6}(a), $R_{d}(V)$ is shown for contacts produced
by consecutive electric short-circuit of the same contact from
$R_{N}=20\,\Omega $ down to $R_{N}=4\,\Omega $. It is seen that as
the contact resistance is reduced, humps start to appear to a
greater degree, and the intensity of the local maximum at zero
bias increases. The latter is possible to explain, proceeding from
\cite{Mazin} where it is shown that in the diffusive regime
$R_{d}(V)$ looks like the presence of a barrier with $Z$= 0.55 in
the ballistic regime. In other words, for a clean S--N boundary
without a barrier ($Z$ = 0) in the diffusive regime, it is
possible to observe a maximum in $R_{d}(V)$ at zero bias, similar
to the ballistic case with $Z=0.55$. In reality, from
Fig.\,\ref{yf6}(c) it is seen that the parameter $Z$ grows with
the reduction of contact resistance, coming nearer to the value
0.55. It is necessary to note that the $\Delta$ value for these
contacts is practically independent of their resistance, remaining
around 2.2 $\pm $ 0.1\,meV. On the other hand, the parameter
$\Gamma $ grows strongly with resistance, as observed
experimentally for many superconductors of different sorts (see
Fig.\,12.14 in \cite{PCSbook}), although the reason for this
remains under discussion.

\section{Discussion}

The presented point contact measurements show a distribution of
the SC gap in the electronic density of states of an
YNi$_{2}$B$_{2}$C film. Important and essential is the fact that
irrespective of the size of the gap, ranging from 1.5\,meV to
2.4\,meV, the observed critical temperature corresponds to the
$T_{\rm c}$ of the film. This excludes the possibility of a
distribution of the gap on the film surface owing to its
non-uniform structure. Another possible source of a gap variation
is the proximity effect resulting in gap suppression at the
surface. However, we assume that in this case the gap value should
correlate inversely with the point contact resistance, in other
words with the size of the contact: the smaller the size, the
lower the gap value measured. However, this is not the case. The
gap value varies even for contacts with nearly equal resistance
[Fig.\,\ref{yf5}(inset)], and has a tendency, if any, to rise with
increasing resistance [Fig.\,\ref{yf6}(b)].

Hence, it is natural to first assume the presence of a gap
anisotropy. Unfortunately, the use of a film does not allow
measurements to be made along different crystallographic
directions. However, as mentioned previously, the film is
predominantly {\it c} axis oriented, and consists of small
crystallites. Thus it is reasonable to expect that with contacts
prepared by touching a film plane with a needle, we will mainly
register a gap along the {\it c} axis. Yet, taking into account
that the contribution to point contact conductivity result from
some finite solid angle, the point contact axis is not well fixed
along the {\it c} axis and with the possible misorientation of
some crystallites our data can be interpreted as the observation
of the gap anisotropy with $\Delta_{\rm max}$/$\Delta_{\rm
min}\approx $ 1.6. We also note that, as seen from
Fig.\,\ref{yf4}, the ratio 2$\Delta_{\rm max}/k_{\rm B}T_{\rm c}$
is close to the standard BCS value of 3.52.

As to the presence of a hybrid gap with $s+g$ wave order parameter
symmetry and point nodes in certain directions, we never observed
gaps less than 1.5\,meV. The results of point contact measurements
in \cite{Pratap}, giving $\Delta\approx$ 0.45\,meV along the {\it
a} axis, give rise to some questions. Firstly, the dependence
$\Delta(T)$ for the $a$ directions looks at least atypical;
secondly, the critical temperature in this direction was found to
be two times less than in the bulk; and thirdly, even for the
maximal gap, 2$\Delta_{\rm max}/k_{\rm B}T_{\rm c}\approx $ 2.86,
much lower than the BCS value. This raises the doubt of whether to
take into account that nickel borocarbides are more likely
superconductors with strong coupling. We note that the single
crystals investigated in \cite{Pratap} had a significantly lower
$T_{\rm c}\approx $ 14.6\,K, in comparison with the value 15.5\,K
usually quoted in the literature. Thus, an insufficient sample
quality is apparently responsible for the reported "anisotropy" of
the SC order parameter in \cite{Pratap}.

Considering the role of disorder in anisotropic superconductors,
nonmagnetic impurity scattering has profound effects on hybrid
$s+g$ wave superconductors with point nodes. It is mentioned in
\cite{Maki2} that a disorder-induced quasiparticle energy gap
opens up even for infinitesimal scattering rates. This may be a
reason why we did not observe any nodal features in the SC gap.
However, impurity scattering also decreases the critical
temperature, which is not observed in our investigations. Thus,
according to \cite{Maki2}, a disorder-induced energy gap of about
10\,K $\approx $ 1\,meV corresponds to a $T_{\rm c}$ reduction to
13\,K.

Concerning the effect of the pressure produced by point-contact
formation on the gap distribution, we expect that under such a
pressure the critical temperature should also vary in the same way
as the gap. However, no such variation is observed, suggesting
that there is no measurable effect due to pressure in this
experiment.

\begin{figure}
\begin{center}
\includegraphics[width=8cm,angle=0]{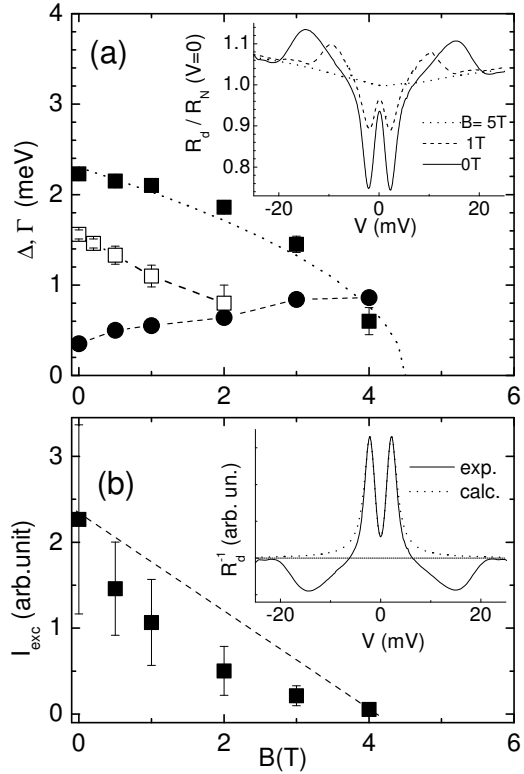}
\end{center}
\caption[] {(a) Calculated magnetic field dependences of $\Delta$
(closed squares) and $\Gamma$ (closed circles) at $Z$ = 0.58 for
an YNi$_{2}$B$_{2}$C--Cu point contact ($R_0$ = 2.8 $\Omega$, $T$
= 4.2\,K). Open squares show the magnetic field behavior of
$\Delta$ for another contact with small gap value. The dotted
curve shows the behavior $\propto(1-B/B_{c2})^{1/2}$ of the pair
potential of a type II superconductor in the vortex state
according to Abrikosov's theory. Dashed lines connect circles and
open squares as guides to the eye. Inset: reduced experimental
$R_d(V)$ in zero field and in magnetic fields of 1\,T and 5\,T.
(b) Behavior of the excess current in a magnetic field. The inset
shows symmetrized experimental and theoretical $R_{d}^{-1}(V)$ in
zero field. The excess current was determined as the integral of
the calculated $R_{d}^{-1}(V)$ after subtraction of the normal
state conductivity (horizontal dashed line). Error bars were
determined as the integral of the difference between theoretical
and experimental $R_{d}^{-1}(V)$ curves. The greatest difference
arises due to the presence of minima in the experimental
$R_{d}^{-1}(V)$ curves at biases exceeding the gap energy. }
\label{yf7}
\end{figure}

Let's consider the question of the realization of multiband
superconductivity in {\it R}Ni$_{2}$B$_{2}$C ({\it R} = Y, Lu),
considered in \cite{Shulga}. We cannot distinguish whether the
observed gap distribution is a consequence of anisotropy or a
multiband electronic structure. In this connection, we note that
the authors of Ref.\,\cite{Bobrov}, investigating point contact
spectra of a LuNi$_{2}$B$_{2}$C single crystal, found that the
calculated theoretical curves better describe the experimental
data when a two-gap adjustment is used, although in the $R_d(V)$
curves, only one gap minimum is visible. In so doing, the relation
between the maximal and minimal gap was found to be around 1.75 in
the {\it c} direction and 1.5 in the $ab$ plane. This is close to
the value found by us, $\Delta_{\rm max}/\Delta_{\rm min}\approx $
1.6. To this it is possible to add that our preliminary data after
analysis of about 60 $R_{d}(V)$ spectra measured on a cleaved
YNi$_2$B$_2$C single crystal also show a gap distribution. In this
case for 80\% of the contacts the gap is distributed in the same
range from 1.5\,meV to 2.4\,meV, while for 20\% of the contacts
the gap ranged more from 2.5\,meV to 3.0\,meV, and one can discern
maxima in the entire gap distribution at 2.0\,meV and 2.4\,meV.
Interestingly, to fit the experimental $dV/dI$ curve for the {\it
c} direction in LuNi$_2$B$_2$C in Ref.\,\cite{Bobrov} (Fig.\,13),
the authors used a gap distribution with two peaks also around
2.0\,meV and 2.5\,meV.

A prominent feature of a multiband superconductor is also the
specific behavior of the excess current in the point contacts. In
\cite{Yans} it was revealed that in the two-band superconductor
MgB$_{2}$, the excess current has a strongly pronounced positive
curvature in a magnetic field. In Fig.\,\ref{yf7}, the behavior of
the SC gap and the excess current in a magnetic field for one of
the YNi$_{2}$B$_{2}$C contacts is shown. It can be seen that,
while the behavior of the gap is similar to that of the pair
potential of a type II superconductor in the vortex state and has
negative curvature [Fig.\,\ref{yf7}(a), closed squares and dotted
curve], the excess current has a strong positive curvature
[Fig.\,\ref{yf7}(b)]. A similar behavior of the gap and the excess
current in a magnetic field was observed by us for several of the
investigated contacts ($R_{d}(V)$ have been processed for four
contacts). Another consideration is that a small gap has an almost
linear decrease under a magnetic field [Fig.\,\ref{yf7}(a), open
squares]. This is similar to the theoretical behavior of the small
gap in MgB$_{2}$ calculated in \cite{Koshelev}. Thus, our
observation of the excess current and small gap behavior under a
magnetic field may be taken in support of a multiband SC state in
YNi$_{2}$B$_{2}$C.

\section{Conclusion}

Studies of the SC gap $\Delta$ and its temperature dependence
$\Delta(T)$ in YNi$_{2}$B$_{2}$C have shown a distribution of
$\Delta$ values from $\Delta_{\rm min}\approx$ 1.5\,meV up to
$\Delta_{\rm max}\approx$ 2.4\,meV, however in all cases with a
BCS-like $\Delta(T)$ dependence. The observed 2$\Delta_{\rm
max}/k_{\rm B}T_{\rm c}\approx $ 3.66 is also close to the BCS
value of 3.52, and the critical temperature in all cases
corresponded to that of the film, thus excluding the possibility
that the observations result from heterogeneity in the properties
of the film. It is most probable that the distribution of the gap
can be attributed to gap anisotropy, or to the multiband nature of
the SC state in YNi$_2$B$_2$C. The positive curvature in the
behavior of the excess current under applied magnetic field
(Fig.\,\ref{yf7}(b)) tends to support the multiband scenario.
Proceeding from the stated results, and also preliminary
investigations of YNi$_2$B$_2$C single crystals, the presence of
point nodes in the SC order parameter of YNi$_2$B$_2$C is
considered to be unlikely. The analogous conclusion is reached
also in a recent work \cite{Igna} on the basis of the observation
of a sudden disappearance of de Haas--van Alphen oscillations at
the transition to the mixed state. Finally, in a recent preprint
\cite{Mukho}, the authors of Ref.\cite{Pratap} have modified their
earlier claims and come to the conclusion from the analysis of the
field variation of their $dV/dI$ spectra that the unconventional
gap function observed in YNi$_2$B$_2$C could originate from
multiband superconductivity.

\section{Acknowledgements}

The support of the U.\,S. Civilian Research and Development
Foundation for the Independent States of the Former Soviet Union
(grant no. UP1-2566-KH-03), the {\it Deutsche
Forschungsgemeinschaft} within SFB 463 "Rare earth transition
metal compounds: structure, magnetism and transport" and of the
National Academy of Sciences of Ukraine are acknowledged.

{}

\end{document}